\documentclass[11pt]{article}

\usepackage[sc]{mathpazo}
\usepackage[mathscr]{eucal}
\usepackage{amsmath,amsfonts,amssymb,amsthm}
\usepackage{caption}
\usepackage{subcaption}
\usepackage{tgpagella}
\usepackage{hyperref}
\usepackage[round]{natbib}
\usepackage{booktabs}
\usepackage{graphicx}
\usepackage{xcolor}
\usepackage[textsize=tiny,shadow]{todonotes}
\usepackage{mathtools}
\theoremstyle{plain}
\newtheorem{theorem}{Theorem}
\newtheorem{lemma}[theorem]{Lemma}
\newtheorem{proposition}[theorem]{Proposition}

\newtheoremstyle{note}{\topsep}{\topsep}{\slshape}{}{\scshape}{}{ }{}
\theoremstyle{note}

\theoremstyle{remark}

\numberwithin{equation}{section}
\numberwithin{theorem}{section}

\newcommand\scM{{\mathscr M}}

\newcommand\mvector{\boldsymbol}

\newcommand\va{\mvector{a}}
\newcommand\vb{\mvector{b}}

\newcommand\vd{\mvector{d}}
\newcommand\ve{\mvector{e}}

\newcommand\vp{\mvector{p}}
\newcommand\vq{\mvector{q}}
\newcommand\vr{\mvector{r}}

\newcommand\vx{\mvector{x}}

\newcommand\vA{\mvector{A}}
\newcommand\vB{\mvector{B}}
\newcommand\vC{\mvector{C}}
\newcommand\vD{\mvector{D}}

\newcommand\vH{\mvector{H}}
\newcommand\vI{\mvector{I}}
\newcommand\vJ{\mvector{J}}

\newcommand\vM{\mvector{M}}

\newcommand\vP{\mvector{P}}
\newcommand\vQ{\mvector{Q}}

\newcommand\veta{\mvector{\eta}}

\newcommand\vrho{\mvector{\rho}}

\newcommand\vzero{\mvector{0}}

\newcommand\field{\mathbb}

\newcommand\R{\field{R}}

\newcommand\C{\field{C}}
\newcommand\Z{\field{Z}}
\newcommand\M{\field{M}}
\newcommand\N{\field{N}}

\newcommand\diag{\operatorname{diag}}

\newcommand\abs[1]{\lvert #1 \rvert}

\newcommand\mtext[1]{\quad\text{#1}\quad}

\newcommand\defset[2]{\left\{{#1}\;\vert \;\; {#2} \,\right\}}

\title{Non-integrability of charged  three-body problem}
\author{
Maria Przybylska \\
Institute of Physics, University of Zielona G\'ora, \\
 Licealna 9, PL-65--417,  Zielona G\'ora, Poland
\\ e-mail: m.przybylska@if.uz.zgora.pl
\\ and \\
Andrzej J. Maciejewski\\
Janusz Gil Institute of Astronomy, University of Zielona G\'ora, \\
ul. Licealna~9, 65-417,  Zielona G\'ora, Poland \\
e-mail: a.maciejewski@ia.uz.zgora.pl
}
\begin{document}
\mathtoolsset{
mathic 
}
\maketitle
\begin{abstract}
We consider the problem of $n$ points with positive masses interacting
pairwise with forces inversely proportional to the distance between them. In
particular, it is the classical gravitational, Coulomb or photo-gravitational  
$n$-body problem. Under this general form of interaction, we investigate the
integrability problem of three bodies.   We show that the system is not integrable except in one case when two among three interaction constants vanish. In our investigation, we used the Morales-Ramis theorem concerning the integrability of a natural Hamiltonian system with a homogeneous potential and its generalization.
 \paragraph*{Declaration}The final published version of the article is available at \textbf{\href{https://doi.org/10.1007/s10569-025-10237-3}{https://doi.org/10.1007/s10569-025-10237-3} }
\end{abstract}
\section{Introduction}  
\label{sec:centr} 
We consider a system of $n$ points with positive masses $m_1$, \dots, $m_n$
moving in a plane. In an inertial frame, their positions are given by the radius
vectors $\vr_1$, \dots, $\vr_n$. Mutual interactions between points are
described by potential 
\begin{equation}
  \label{eq:Vn}
  V(\vr) = - \sum_{1\leq i< j\leq n}\frac{\gamma_{ij}}{\abs{\vr_i - \vr_j}},
\end{equation}
where $\gamma_{ij}=\gamma_{ji}$ are real constants, and $\vr=(\vr_1, \ldots,
\vr_n)\in (\R^2)^n$. The coefficients $\gamma_{ij}$ for purely gravitational
interactions are $\gamma_{ij}= m_i m_j$, for purely electrostatic interactions equal $
\gamma_{ij}= -e_ie_j$, where $e_i$ are the charges of the bodies and if both of
these interactions have to be taken into account, then $\gamma_{ij}= m_i m_j -
e_ie_j$. 

This problem is called the charged $n$-body problem and can be considered as a
natural generalization of the gravitational $n$-body problem. The charged
two-body problem is integrable for an arbitrary value of  $\gamma_{12}$, see e.g.
\cite[Sec 2]{Bengochea:15::}. It seems that the first extended study of the charged three-body problem in the general formulation was carried out by
\cite{Crater:78::} and \cite[Sec. 2]{Dionysiou:81::}.  Later various properties
of this system were analyzed in articles:
\cite{Atela:88::,Atela:94::,Perez-Chavela:96::,Alfaro:08::,Castro-Ortega:12::,Mansilla:12::,Castro-Ortega:14::,Zhou:15::,Castro-Ortega:16::,Llibre:17::,Zaman:2017::,Hoveijn:19::,Hoveijn:23::,Hoveijn:23::n}.
The  charged versions of restricted three and more bodies  problems were studied by
\cite{Dionysiou:87::,Dionysiou:89::,Llibre:13::,Bengochea:15::,Palacian:18::,Perez-Rothen:22::},
and by
\cite{Casasayas:90::,Casasayas:91::,Afaro:00::,Alfaro:02::,Alfaro:02::a,Hoveijn:23::n}, respectively.
The relativistic version of the problem was also considered e.g. by \cite{Barker:77::,Dionysiou:81::}.

The system is Hamiltonian, and its  Hamiltonian function written  in canonical variables $\vr_i$ and $\vp_i = m_i\dot\vr_i$  reads 
\begin{equation}
  \label{eq:ham}
  H = \frac{1}{2}\vp^T \vM^{-1}\vp + V(\vr),
\end{equation}
where $\vp=(\vp_1, \ldots,
\vp_n)$, and 
\begin{equation}
  \vM=\diag(m_1,m_1, \ldots, m_n, m_n)=
  \begin{bmatrix}
    m_1 \vI_2 & \vzero & \ldots & \vzero \\
    \vzero & m_2 \vI_2 & \ldots & \vzero \\
    \hdotsfor{4}\\
    \vzero &\vzero & \ldots  &m_n \vI_2
  \end{bmatrix},
\end{equation}
and $\vI_2$ is $2\times2$ identity matrix. We use matrix notation; a vector
$\vx\in\R^k$ is considered as a matrix of one column, so $\vx=[x_1, \ldots,
x_k]^T$. Moreover, we introduce vector $\vr=(\vr_1, \ldots, \vr_n)=[\vr_1^{T}, \ldots,
\vr_n^T]^{T} $.  For two vectors $\va, \vb\in \R^k$, we also use the notation $\va\cdot\vb=\va^T\vb$. 

The total linear and angular momenta  
\begin{equation}
  \label{eq:momenta}
  \vP=\sum_{i=1}^n \vp_i, \qquad C= \sum_{i=1}^n \vr_i^T\vJ_2^T\vp_i,
  \qquad
  \vJ_2 =
  \begin{bmatrix}
  0 & -1 \\
  1& 0
  \end{bmatrix}, 
\end{equation}
are the first integrals of the system. Notice, that they do not commute 
\begin{equation}
  \{C,P_1\}=P_2\mtext{and} 
  \{C, P_2\} = -P_1.
  \end{equation}
The system has $2n$ degrees of freedom and for its integrability in the
Liouville sense $2n$ functionally independent and pairwise commuting first
integrals are needed. In our further investigations, we consider the system in the center of the mass frame.

The aim of this article is to study the integrability of the system.  The main result of this
paper concerns the three-body problem. For this case we write the potential in
the following form 
\begin{equation}
  \label{eq:V3}
  V(\vr)= -\frac{\alpha_3}{\abs{\vr_1 - \vr_2}}
  -\frac{\alpha_2}{\abs{\vr_2 - \vr_3}} 
  -\frac{\alpha_1}{\abs{\vr_3 - \vr_1}}.
\end{equation}
Our main result is the following theorem. 
\begin{theorem}
\label{thm:main} If at least two of the parameters $\alpha_1$, $\alpha_2$ and
$\alpha_3$ are non-zero, then the charged three-body problem is not integrable
in the Liouville sense. 
\end{theorem}  
In order to prove this result we will use a differential Galois approach to the integrability. In the context of Hamiltonian systems the main theorem of this
approach is the following. 
\begin{theorem}[Morales,1999]
  \label{thm:mo}
If a Hamiltonian system is integrable with complex meromorphic first integrals
in the Liouville sense, then the
identity component of the differential Galois group of the  variational
equations along a particular solution is Abelian.
\end{theorem}
Numerous successful applications of the above theorem show that it offers a very powerful and effective tool for studying the integrability see e.g. \cite{Morales:10::} and references therein.  The considered  
system is described by a natural Hamiltonian and the potential is a homogeneous function of degree $k=-1$. For systems of such a form  from
Theorem~\ref{thm:mo} one can deduce particularly efficient criteria for the
integrability.   We explain this in the next section.

In order to apply Theorem~\ref{thm:mo} for the considered system we have to
extend it to the complex phase space. Thus, we will assume that $\vr_i\in\C^2$
and $\vp_i\in\C^2$ for $i=1, \ldots, n$. We will keep the same notation for
real and complex vectors but  for a complex $\va \in\C^k$  we denote 
\begin{equation}
  \abs{\va}:=\left( \sum_{i=1}^ka_i^2 \right)^{1/2}. 
\end{equation} 
Notice that the potential~\eqref{eq:Vn} is not a complex
meromorphic function. Therefore,  we cannot apply directly Theorem~\ref{thm:mo}
for investigation of the integrability of the system given by Hamiltonian~\eqref{eq:ham}. However, $V(\vr)$ is
an algebraic function, so using the method described by~\cite{Combot047475} one can
extend the application of Theorem~\ref{thm:mo}  to
systems with such potentials, see also \citep{mp:16::a}.  
Shortly speaking, we consider the integrability with first integrals which are meromorphic functions of variables $(\vr,\vp,\vrho)\in(\C^2)^n\times(\C^2)^n\times\left(\C^{\ast}\right)^{n(n-1)/2}$, where $\vrho$ is $n(n-1)/2$ dimensional vector with coordinates $\rho_{ij}=|\vr_i-\vr_j|$ for $1\leq i<j\leq n$.

\section{Non-integrability criterion}
For an application of Theorem~\ref{thm:mo} one needs a particular solution.
Unfortunately,  finding such a solution is difficult. However, if the
considered system is natural and the potential function is homogeneous, then we
have a procedure how to find them. It is described in
\cite{Yoshida:87::a}, see also \cite{mp:09::a}. Here we present it with a modification
convenient for studying many body systems.

Let us consider a natural complex  Hamiltonian system with $n$ degrees of
freedom described by a Hamilton function of the following form
\begin{equation}
  \label{eq:hamm}
  H = \frac{1}{2}\vp^T\vM^{-1}\vp +V(\vq),          
\end{equation}
where $\vq=(q_1,\ldots,q_n)\in\C^n$ and $\vp=(p_1,\ldots,p_n)\in\C^n$ are
canonical coordinates and momenta, potential $V(\vq)$ is a meromorphic
homogeneous function of degree $k\in\Z^{\star}$, and $\vM$ is a symmetric
non-singular $n\times n$ matrix.  For such a system with $\vM=\vI_n$, simple and
very strong conditions for integrability were formulated in the book
\cite{Morales:99::c}. Similar conditions can be easily found in a more general
case when $\vM\neq \vI_n$, see \citep{mp:11::a}.

The basic assumption is that there exists a nonzero vector $\vd\in\C^{n}$ such
that
\begin{equation}
  \label{eq:dar}
 \nabla V(\vd)=\mu \vM \vd,
\end{equation}
where $\mu\in\C$ is a non-zero constant and $ \nabla V(\vq)$ denotes the gradient of $V(\vq)$. Such a vector is 
 called a Darboux point of
potential $V(\vd)$ and it defines a particular solution of the system.
In fact, $(\vq(t), \vp(t))=(x(t) \vd, \dot x(t) \vM\vd)$ is a solution of
Hamilton's equations generated by~\eqref{eq:hamm} provided that $x(t)$ is a
solution of Newton's equation
\begin{equation}
  \label{eq:scalar}
  \ddot x = -\mu x^{k-1}.
\end{equation}
 The variational equations along this
particular solution can be written in the form 
\begin{equation}
  \label{eq:var2}
  \ddot \vQ = -\mu x(t)^{k-2}\vH(\vd)\vQ, 
\end{equation}  
where 
\begin{equation}
  \label{eq:hess}
  \vH(\vd) = \mu^{-1} \vM^{-1} \nabla^2 V(\vd)
\end{equation}
and $\nabla^2 V(\vq)$ is the Hessian matrix of potential $V(\vq)$.
Let us make a linear change of variables $\vQ=\vA\veta$,  which transforms the
above system to the following one
\begin{equation}
  \label{eq:jordan}
  \ddot \veta =  - \mu x(t)^{k-2} \vB \veta,
\end{equation}
where  $\vB$ is the Jordan form of  matrix $\vH(\vd)$. That is, $\vB$ has a
block diagonal form 
\begin{equation}
  \label{eq:jor}
  \vB = \vA^{-1}\vH(\vd)\vA = \diag(\vb(\lambda_1, n_1), 
  \ldots \vb(\lambda_p, n_p) ),
\end{equation}
where  $\lambda_i$ are   eigenvalues of  $\vH(\vd)$ and 
\begin{equation}
  \vb(\lambda, m)= \begin{bmatrix}
    \lambda & 0 &0& \hdotsfor{2}& 0 \\
    1& \lambda & 0& \hdotsfor{2}&0\\
    0& 1&\lambda& \hdotsfor{2}&0 \\
    \hdotsfor{6}\\
    0& 0& 0& \ldots &1& \lambda
  \end{bmatrix}
  \in\M(m,\C)
\end{equation}
and   $\M(m,\C)$ denotes the set of $m\times m$ complex matrices. 
The differential Galois group of equation~\eqref{eq:jordan} can be investigated
effectively thanks to Haruo Yoshida's brilliant transformation
\begin{equation}
  \label{eq:yosan}
  t \longmapsto z = \frac{\mu}{ke}x(t)^2, \qquad e = \frac{1}{2}\dot x(t)^2 +\frac{\mu}{k}x(t)^k.
\end{equation}
After this transformation \eqref{eq:jordan} reads 
\begin{equation}
  \label{eq:afteryo}
  2kz(1-z)\veta''  +\left[ z(2-3k)+ 2(k-1)\right]\veta' = -\vB \veta.
\end{equation}
This equation  splits into subsystems corresponding to  Jordan blocks and each
block contains a scalar equation of the form 
\begin{equation}
  \label{eq:sc2}
  2kz(1-z)\eta''  + \left[ z(2-3k)+ 2(k-1)\right]\eta' = -\lambda\eta,
\end{equation} 
where $\lambda$ is an eigenvalue of  $\vH(\vd)$. This is the Gauss
hypergeometric equation. Using Kimura theorem \citep{Kimura:69::} one can easily
find values of $(k,\lambda)$ for which the identity component of the differential Galois group of this
equation is solvable. Then by Theorem~\ref{thm:mo} one can formulate necessary integrability conditions for natural Hamilton systems with homogeneous potentials which was made in \cite{Morales:01::a,Morales:99::c}.  If matrix $\vB$ is not diagonal, then one can find additional integrability obstructions. However, analysis of these
cases is quite involved, see \citep{Maciejewski:09::}. Collecting all these obstructions to
the integrability one can formulate the following theorem. 
\begin{theorem}
  \label{thm:mo2}
  Assume that the Hamiltonian system defined by the Hamiltonian
  \eqref{eq:hamm} with a homogeneous potential $V(\vq)$ of degree
  $k\in\Z^\star:=\Z\setminus\{0\}$ satisfies the following conditions:
  \begin{enumerate}
  \item there exists a non-zero $\vd\in\C^n$ such that $
    \nabla V(\vd)=\mu\vM\vd$, and
  \item the system is integrable in the Liouville sense with  meromorphic first
  integrals.
  \end{enumerate}
  Then for each eigenvalue $\lambda$ of matrix $\vH(\vd)$, pair
  $(k,\lambda)$ belongs to an item of the following table
  \begin{equation}
    \label{eq:tab}
    \begin{array}{crcr}
      \toprule
      \text{case} &  k & \lambda  &\\
      \midrule 
      \vbox to 1.3em{}1. &  \pm 2 &  \text{arbitrary} & \\
      \addlinespace[0.5em]
      2.  & k & p + \dfrac{k}{2}p(p-1) & \\
      \addlinespace[0.5em]
      3. & k & \dfrac 1 {2}\left(\dfrac {k-1} {k}+p(p+1)k\right)  & \\
      \addlinespace[0.5em]
      4. & 3 &  -\dfrac 1 {24}+\dfrac 1 {6}\left( 1 +3p\right)^2, & -\dfrac 1
      {24}+\dfrac 3 {32}\left(  1  +4p\right)^2 \\
      \addlinespace[0.5em]
      & & -\dfrac 1 {24}+\dfrac 3 {50}\left(  1  +5p\right)^2,  &
      -\dfrac1{24}+\dfrac{3}{50}\left(2 +5p\right)^2 \\
      \addlinespace[0.5em]
      5. & 4 & -\dfrac 1 8 +\dfrac{2}{9} \left( 1+ 3p\right)^2 & \\
      \addlinespace[0.5em]
      6. & 5 & -\dfrac 9 {40}+\dfrac 5 {18}\left(1+ 3p\right)^2, & -\dfrac 9
      {40}+\dfrac 1 {10}\left(2
        +5p\right)^2\\
      \addlinespace[0.5em]
      7. & -3 &\dfrac {25} {24}-\dfrac 1 {6}\left( 1 +3p\right)^2, & \dfrac {25}
      {24}-\dfrac 3 {32}\left(1 +4p\right)^2 \\
      \addlinespace[0.5em]
      & & \dfrac {25} {24}-\dfrac 3 {50}\left(1+ 5p\right)^2, & \dfrac {25}
      {24}-\dfrac{3}{50}\left(2+ 5p\right)^2\\
      \addlinespace[0.5em]
      8. & -4 & \dfrac 9 8-\dfrac{2}{9}\left( 1+ 3p\right)^2 & \\
      \addlinespace[0.5em]
      9. & -5 & \dfrac {49} {40}-\dfrac {5} {18}\left(1+3p\right)^2 ,& \dfrac {49}
      {40}-\dfrac {1} {10}(2 +5p)^2\\
      \addlinespace[0.5em]
      \bottomrule
    \end{array}
  \end{equation}
where $p$ is an integer. Moreover, for $k\not\in\{-2,0,2\}$, 
\begin{itemize}
  \item matrix $\vH(\vd)$ does not have a Jordan block of size
    $d\geq 3$;
  \item if matrix $\vH(\vd)$ has a Jordan block 
    size $d=2$, then the corresponding eigenvalue
    $\lambda$ satisfies the following
    conditions:
    \begin{enumerate}
    \item if $\abs{k}>2$, then $\lambda$ does not belong to the second
      item of table~\eqref{eq:tab};
    \item if $k=-1$, then $\lambda=1$;
    \item if $k=1$, then $\lambda=0$.
    \end{enumerate}
\end{itemize}
\end{theorem}
\section{Central configurations and integrability conditions}
\label{ssec:central}

Points with masses $m_i$ and positions $\vd_i$, $i=1, \ldots, n$  form a central
configuration  $ \vd= (\vd_1, \ldots, \vd_n)\in \left( \R^2\right)^n$ if 
\begin{equation}
\label{eq:cc_exp}
\sideset{}{'}\sum_{j=1}^n \frac{\gamma_{ij}}{\abs{\vd_i - \vd_j}^3}\left( \vd_i -
\vd_j\right)
= \mu m_i \vd_i
, \qquad i = 1, \ldots,n,
\end{equation}
for a certain $\mu\in\R$. The primed sum sign denotes the summation over $j\neq
i$. We can rewrite these equations in the vector form using the gradient of potential \eqref{eq:Vn} as
\begin{equation}
\label{eq:cc}
\nabla V(\vd) = \mu \vM\vd.
\end{equation}
Thus, in the terminology from the previous section, a central configuration is a Darboux point of potential $V(\vr)$.

It is easy to show that
\begin{equation}
\label{eq:lam}
\mu := \mu(\vd)=-\frac{V(\vd)}{I(\vd)},  \qquad I(\vd)= \vd^T \vM \vd.
\end{equation}
Notice that if $\vd$ is a central configuration, then, for an arbitrary
$\vA\in\mathrm{SO}(2,\R)$, $\widetilde\vd:=\widehat{\vA}\vd:=(\vA\vd_1, \ldots,
\vA\vd_n)$ is also a central configuration with the same $\mu$. Moreover, if
$\vd$ is a central configuration, then, $\widetilde\vd=\alpha\vd$ is also a
central configuration for an arbitrary $\alpha\neq 0$, with $\mu(
\widetilde\vd)= \alpha^{-3} \mu(\vd)$.  Therefore, central configurations are
not isolated.  Let us notice that in Newtonian $n$-body problem $\mu>0$ but for
charged $n$-body problem $\mu\in\R$.  More details about central configurations can be found in \cite{MR3469182}.

\begin{proposition}
  Assume that $ \vd= (\vd_1, \ldots, \vd_n)$  is a central configuration. Then 
  $\vr(t)= x(t)\vd$ is a solution of Newton's equation 
  \begin{equation}
    \label{eq:newton}
    \vM\ddot\vr = -\nabla V(\vr),
  \end{equation}
  provided $x(t)$ is a solution of one dimensional Kepler problem 
  \begin{equation}
    \label{eq:kepler1}
    \ddot x = -\frac{\mu(\vd)}{x^2}.
  \end{equation}
\end{proposition}
A central configuration gives a more general solution, 
which is called the Kepler homographic solution: each body moves along a Keplerian orbit around the center of mass of the system, see~\cite{MR3469182}.

For a collinear configuration $ \vd= (\vd_1, \ldots, \vd_n)$ points $ \vd_i$ lie
on a straight line. Without loss of the generality, we can assume that in this case $\vd_i =(x_i,0)$ for $i=1, \ldots, n$.   Let us assume that such a central configuration exists.  To apply Theorem~\ref{thm:mo2} we have to calculate matrix $\vH(\vd)$ defined in~\eqref{eq:hess}.

Matrix $\vH(\vr)=\mu^{-1}\vM^{-1}\nabla^2 V(\vr)$ has the following block
structure 
\begin{equation}
\label{eq:H(r)}
\vH(\vr) =
\begin{bmatrix}
\vH_{11}(\vr) & \ldots & \vH_{1n}(\vr)  \\
\hdotsfor{3} \\
\vH_{n1}(\vr) & \ldots & \vH_{nn}(\vr)  \\
\end{bmatrix},
\end{equation}
where $2\times2$ blocks $\vH_{ij}(\vr)$ for $i\neq j$ are given by
\begin{equation}
\label{eq:Hij} 
\vH_{ij}(\vr)= -\frac{\lambda_{ij}}{\mu m_i\abs{\vr_i - \vr_j}^3}
\left[ \vI_2 - 3 \frac{(\vr_i-\vr_j)(\vr_i-\vr_j)^T}{\abs{\vr_i - \vr_j}^2}
\right],
\end{equation}
and for $j=i$ are defined as 
\begin{equation}
\label{eq:Hii}
\vH_{ii} (\vr)= -\sideset{}{'}\sum_{j=1}^n \vH_{ij}(\vr).
\end{equation}
From the above formulae, we can easily deduce that for an arbitrary $\va\in\R^2$ 
$\vH(\vr)(\va, \ldots, \va)=\vzero$.  
\begin{proposition}
	\label{pro:Heig}
  Let $ \vd= (\vd_1, \ldots, \vd_n)$ be a central configuration. Then 
	vectors $\vd$ and $\widehat{\vJ}\vd =(\vJ_2 \vd_1, \ldots, \vJ_2\vd_n) $ are eigenvectors of the matrix
	${\vH}(\vd)$ with eigenvalues $-2$ and $1$, respectively.
\end{proposition}
\begin{proof}
	Components of gradient $\nabla V(\vr)$ are homogeneous functions of degree $-2$.
	Thus, by the Euler identity
	\begin{equation}
	\nabla^2V(\vr)\vr = -2 \nabla V(\vr).
	\end{equation}
	Evaluating both sides of this identity at $\vr=\vd$ we obtain 
	\begin{equation}
	\nabla^2V(\vd)\vd =   -2 \nabla V(\vd) = -2 \mu\vM \vd 
	\end{equation}
	and thus 
	\begin{equation}
	\vH(\vd)\vd = \mu^{-1}\vM^{-1}\nabla^2V(\vd)\vd = -2 \vd.
	\end{equation}
	We already mentioned that if $\vd$ is a central configuration, then $\widehat{\vA}\vd$ is
	also a central configuration for arbitrary $\vA\in\mathrm{SO}(2,\R)$. Therefore, we have the following equality 
	\begin{equation}
	\nabla V( \widehat{\vA}(\theta)\vd) = \mu \vM \widehat{\vA}(\theta)\vd, \qquad \vA(\theta) = 
	\begin{bmatrix}
	\cos \theta & -\sin \theta \\
	\sin \theta &  \cos \theta
	\end{bmatrix}.
	\end{equation}
	Differentiating both sides of this equality at $\theta=0$ we get 
	\begin{equation}
	\nabla^2 V( \vd)\widehat{\vJ}\vd = \mu \vM \widehat{\vJ}\vd.
	\end{equation}
	Hence,
	\begin{equation}
	\widetilde{\vH}(\vd)\widehat{\vJ}\vd =\mu^{-1}\vM^{-1}\nabla^2 V( \vd)\widehat{\vJ}\vd= \widehat{\vJ}\vd.
	\end{equation}
\end{proof}

For a  collinear central configuration $ \vd= (\vd_1, \ldots, \vd_n)$,  $\vd_i =(x_i,0)$ for
$i=1, \ldots,n$, we have 
\begin{equation}
  \label{eq:Hijs}
    \vH_{ij}(\vd)= C_{ij} \vD, \qquad \vD = 
   \begin{bmatrix}
   -2  & 0 \\
   0  & 1
   \end{bmatrix},
 \end{equation}
 for $i\neq j$
 \begin{equation}
 C_{ij} = -\frac{\lambda_{ij}}{\mu m_i\abs{x_i - x_j}^3},
 \end{equation}
 and
 \begin{equation}
 C_{ii} = -\sideset{}{'}\sum_{j=1}^nC_{ij}.
\end{equation}
Thus, after permutation  
\begin{equation*}
  [x_1, y_1, \ldots, x_n, y_n]^T = \vP [x_1, \ldots, x_n, y_1, \ldots, y_n]^T,
\end{equation*}
matrix  $\vH(\vd)$ get a block diagonal form  
\begin{equation}
  \label{eq:simil}
  \vP\vH(\vd)\vP^{-1}= \begin{bmatrix}
    -2 \vC  & \vzero \\
    \vzero & +\vC
  \end{bmatrix}. 
\end{equation}
Thus, we have the following.
\begin{proposition}
  \label{pro:eig}
If $\ve =(e_1, \ldots, e_n)$ is an eigenvector of matrix $\vC$ corresponding to
eigenvalue $\lambda$, then $(e_1,0,\ldots,e_n,0)$ is an eigenvector of
$\vH(\vd)$ corresponding to eigenvalue $-2\lambda$, and $(0,e_1,\ldots,0,e_n)$
is an eigenvector of $\vH(\vd)$ corresponding to eigenvalue $\lambda$. Moreover,
matrix $\vC$ has at least one eigenvalue $0$ and at least one eigenvalue $1$.
\end{proposition}
With the above facts, we can prove the following theorem. 
\begin{theorem}
  \label{thm:we0}
Let $\vd$ be a collinear central configuration of the charged $n$-body problem. 
If this problem is integrable in the Liouville sense, then all non-zero
eigenvalues of matrix $\vC$ are equal to 1 and this matrix is diagonalizable. 
\end{theorem}
\begin{proof}
The potential $V(\vr)$ is a homogeneous function of degree $k=-1$. We apply
Theorem~\ref{thm:mo2} taking as a Darboux point an arbitrary collinear central
configuration $\vd$. For $k=-1$ only  items 2 and 3 in
table~\eqref{eq:tab} are  admissible  but both of
them define the same set of integers 
\begin{equation}
  \label{eq:M1} 
\scM_{-1}= \defset{\frac{1}{2}(2+p-p^2)}{ p\in \N}=\{ 1,0, -2, -5, \ldots\}.
\end{equation}
By Proposition~\ref{pro:eig} if $\lambda$ is an eigenvalue of $\vC$, then
$\lambda$ and  $-2\lambda$ are eigenvalue of $\vH(\vd)$. Moreover, the Jordan block
corresponding to $\lambda$ has the same dimensions as this corresponding to
$-2\lambda$. Therefore, if the system is
integrable, then $\lambda, -2\lambda\in \scM_{-1}$, and this implies that either 
$\lambda=0$, or  $\lambda=1$. By Theorem~\ref{thm:mo2} Jordan blocks of
$\vH(\vd)$ has dimension smaller than 3. If matrix $\vC$ has a two-dimensional
block, then matrix  $\vH(\vd)$ has a pair of blocks of the same dimension: one
corresponding to $\lambda$ and the second $-2\lambda$. For $\lambda\in\{0,1\}$
the existence of such blocks implies non-integrability.
\end{proof}
F.\,R.~Moulton in \cite{Moulton:1910::} proved that in the classical $n$-body
problem,   for every ordering of $n$ positive masses, there exists a unique
collinear central configuration. There is no similar statement for the charged $n$-body problem. Probably it is valid in the case when all parameters
$\gamma_{ij}$ have the same sign. In general, the knowledge concerning central
configurations in the charged $n$-body problem is very limited. Complete
investigations of central configurations in the char\-ged three-body problem were presented in \citep{Perez-Chavela:96::,Alfaro:08::}. It was shown that
there exist, as in the classical three-body problem, collinear and triangular
central configurations. The triangular central configurations exist only when
all $\gamma_{ij}$ have the same sign. Moreover, the shape of the triangle in the
triangular configuration can be arbitrary. That is,  for a given triangle, there
exists a choice of $\gamma_{ij}$ such that the masses located at the vertices of
this triangle form a central configuration. If not all $\gamma_{ij}$ vanish, then
there exists at least one collinear central configuration. Moreover,  their number can vary
from one to five.

\section{Proof of Theorem~\ref{thm:main}}

Our proof is based on the criterion given by Theorem~\ref{thm:we0}.  
Thus, we will investigate variational equations along a solution corresponding to
the collinear central configuration. In fact, it reduces to study of eigenvalues of the Hessian matrix of the potential evaluated at the central configuration.
The proof splits into three steps. In the first step, we use the fact that there
exists at least one collinear real central configuration. From the conditions for the integrability we deduce among other things, that if the system is integrable,
then there exist at least two collinear configurations. In the next step, using
the second configuration we found additional conditions. Combining both of them
we prove that the system is not integrable under generic assumption
$\alpha_1\alpha_2\alpha_3 \neq 0$.  In the last step we investigate a case when among
parameters $(\alpha_1\alpha_2\alpha_3)$ one vanishes and the other two are
different from zero. 

\begin{lemma}
  \label{lem:l1}
If $\alpha_1\alpha_2\alpha_3 \neq 0$, then the charged three-body problem is not
integrable in the Liouville sense. 
\end{lemma}
\begin{proof}

We assume $\alpha_1\alpha_2\alpha_3 \neq 0$. In this case, we know that there
exists at least one collinear central configuration $\vd=(\vd_1,\vd_2, \vd_3)$,
with $\vd_i\in\R^2$, 
see   Theorem 7.1 in \citep{Perez-Chavela:96::}. We can assume that
$\vd_i=(x_i,0)$ for $i=1,2,3$. Moreover, we assume that masses are located at
$x$ axis at order $(1,2,3)$, so we number points in such a way that $x=x_2-x_1>0$
and $y=x_3-x_2>0$, and $z =x_3-x_1=x+y>0$. With this parametrization equations
defining the central configuration~\eqref{eq:cc_exp} read 
\begin{equation}
  \label{eq:ccc3}
  \left\{\quad
  \begin{aligned}
    -\frac{\alpha _1}{z^2}-\frac{\alpha _3}{x^2}  =  &\mu  m_1 x_1 ,\\
  -\frac{\alpha _2}{y^2}+\frac{\alpha
   _3}{x^2}= &\mu  m_2 x_2,\\
   \frac{\alpha _1}{z^2}+\frac{\alpha _2}{y^2}= &\mu  m_3 x_3.
  \end{aligned}
  \right.
\end{equation} 
Now, we take linear combinations of these equations with coefficients $(m_2,
-m_1, 0)$, $(0,m_3, -m_2)$ and $(-m_3, 0, m_1)$.  In effect, we obtain 
\begin{equation}
  \label{eq:ccc3xyz}
  \left\{\quad
  \begin{aligned}
    \frac{\alpha _3 \left(m_1+m_2\right)}{x^2}-\frac{\alpha _2 m_1}{y^2}+\frac{\alpha _1 m_2}{z^2} = & \mu  m_1 m_2 x, \\
    -\frac{\alpha _3 m_3}{x^2}+\frac{\alpha _2 \left(m_2+m_3\right)}{y^2}+\frac{\alpha _1 m_2}{z^2} = & \mu  m_2 m_3  y,  \\
    \frac{\alpha _3 m_3}{x^2}+\frac{\alpha _2 m_1}{y^2}+\frac{\alpha _1 \left(m_1+m_3\right)}{z^2}= & \mu  m_1 m_3 z. 
  \end{aligned}
  \right.
\end{equation} 
As $z = x+y$, it is convenient to introduce variable $u=y/x$, $z=x(1+u)$  and
then eliminating $\mu$ from the above system we get a single polynomial equation for $u$ of the following form
\begin{multline}
  \label{eq:polu}
  m_2 m_3 u^3 \left[\alpha _1+\alpha _3 (u+1)^2\right]+ 
  m_1 m_3 (u+1)^3 \left(\alpha _3 u^2-\alpha _2\right)\\-m_1 m_2
  \left[\alpha _1 u^2+\alpha _2 (u+1)^2\right]=0.
\end{multline}
A positive root of this polynomial defines the central configuration.

Matrix $\vC$ is given by 
\begin{equation}
  \vC = \frac{1}{\mu z^3}\widetilde{\vC},
\end{equation}
where 
\begin{equation}
  \widetilde{\vC} = \begin{bmatrix}
    \dfrac{1}{m_1} & 0 & 0 \\
    0 & \dfrac{1}{m_2} & 0 \\
    0 & 0 & \dfrac{1}{m_3} \\
  \end{bmatrix}
  \begin{bmatrix}
    \alpha _3+\dfrac{\alpha _1}{(u+1)^3} & -\alpha _3 & -\dfrac{\alpha _1}{(u+1)^3} \\[1em]
    -\alpha _3 & \alpha _3+\dfrac{\alpha _2}{u^3} & -\dfrac{\alpha _2}{u^3} \\[1em]
    -\dfrac{\alpha _1}{(u+1)^3} & -\dfrac{\alpha _2}{u^3} & \dfrac{\alpha _2}{u^3}+\dfrac{\alpha _1}{(u+1)^3} 
  \end{bmatrix}.
\end{equation}
We consider the eigenvalue problem for matrix $\widetilde{\vC}$. We know that
its one eigenvalue is zero. If the system is integrable, then the remaining two
coincide and do not vanish. The problem is that we do not know the roots of
polynomial~\eqref{eq:polu}. However, it depends linearly on the parameters
$\alpha_i$. Thus, solving~\eqref{eq:polu} for $\alpha_2$ we find 
\begin{equation}
  \label{eq:a2}
  \alpha _2 =u^2 \frac{\alpha _3 m_3 (u+1)^2 \left[m_1 (u+1)+m_2 u\right]+\alpha _1 m_2 \left(m_3
   u-m_1\right)}{m_1 (u+1)^2 \left[m_3 (u+1)+m_2\right]}.
\end{equation} 
After substitution this expression  into matrix $\widetilde{\vC}$ we find
directly that its characteristic polynomial factors and two nonzero eigenvalues
are 
\begin{equation}
  \label{eq:lambdy}
  \begin{split}
    \lambda_1 = & \frac{\left(m_1+m_2+m_3\right) \left[\alpha
    _1+\alpha _3 (u+1)^2\right]}{m_1 (u+1)^2 \left[m_3
    (u+1)+m_2\right]},\\
  \lambda_2 = & \frac{\alpha _1+\alpha _3
  (u+1)^3}{m_1(u+1)^3}+\frac{\alpha _3 m_3 (u+1)^4-\alpha
  _1 m_2}{m_2 m_3 u(u+1)^3}.
  \end{split}
\end{equation}
Equation $\lambda_1=\lambda_2$ has two solutions
\begin{equation}
  \label{eq:sol}
  \alpha_1 = \frac{\alpha_3 m_3 (u+1)^3}{m_2},
   \qquad  m_2=-\frac{m_1 m_3 (u+1)^2}{m_3u^2+m_1}.
\end{equation}
The second solution is excluded because all masses are positive.  
With  the first solution from~\eqref{eq:a2} we get 
\begin{equation}
  \label{eq:a2so}
  \alpha _2= \frac{\alpha _3 m_3
   u^3}{m_1}.
\end{equation}
Notice that by definition $u>0$, therefore if the necessary condition for
integrability is fulfilled, then parameters $\alpha_i$ have the same signs. 

It is convenient to introduce two positive  parameters 
\begin{equation}
  \label{eq:gamma12}
   \gamma_1 =\frac{m_1\alpha_2}{m_3 \alpha_3}, \qquad 
   \gamma_2 =\frac{m_2\alpha_1}{m_3 \alpha_3}.
\end{equation} 
Now, we can conclude our reasoning in the following way. If the system is
integrable, the parameters $(\gamma_1,\gamma_2)$ belongs to the curve $\Gamma_1$
given parametrically by 
\begin{equation}
  \gamma_1 = u^3, \qquad \gamma_2 = (1+u)^3, \qquad u>0.
\end{equation}

We already remarked that if the necessary condition for integrability is
fulfilled, then all $\alpha_i$ have the same sign. Thus, using  Theorem 7.1 from  
\citep{Perez-Chavela:96::}, we know that there exist three collinear central
configurations as in the gravitational three-body problem. For the  second  central
configuration we assume that masses $m_i$ are located on $x$ axis at order
$(2,1,3)$, Then, $x_1 -x_2 =x>0$, $x_3 -x_1 =y>0$   and $x_3 -x_2 =z=x+y>0$. 
Proceeding as for the first collinear central
configuration we find out that the second central configuration corresponds to a positive root of the polynomial   
\begin{multline}
  \label{eq:elu2}
  m_2 m_3 (v+1)^3 \left(\alpha _3 v^2-\alpha _1\right)+m_1 m_3 v^3 
  \left(\alpha _2+\alpha _3 (v+1)^2\right)\\-m_1m_2
   \left(\alpha _2 v^2+\alpha _1 (v+1)^2\right)=0,
\end{multline}
where $v=y/x$.
Now, from this equation we determine $\alpha_1$ 
\begin{equation}
  \label{eq:al1}
  \alpha _1 = v^2 \,\frac{\alpha _3 m_3 (v+1)^2 \left[m_1 v+m_2 (v+1)\right]+\alpha _2 m_1 \left(m_3
   v-m_2\right)}{m_2 (v+1)^2 \left[m_3 (v+1)+m_1\right]}.
\end{equation}
With this $\alpha_1$ the characteristic polynomial of matrix $\widetilde{\vC}$
factors and its two nonzero eigenvalues
are 
\begin{equation}
  \label{eq:lambdy2}
  \begin{split}
    \lambda_1 = & \frac{\left(m_1+m_2+m_3\right) 
    \left(\alpha _2+\alpha _3 (v+1)^2\right)}{m_2 (v+1)^2 \left(m_3
    (v+1)+m_1\right)},\\
  \lambda_2 = & -\frac{\alpha _2}{m_3v (v+1)^3}+
  \frac{\alpha _3 (v+1)}{m_1v}+\frac{v \left(\alpha _2+\alpha _3
  (v+1)^3\right)}{m_2v (v+1)^3}.
  \end{split}
\end{equation}
Now equation $\lambda_1=\lambda_2$ has two solutions
\begin{equation}
  \label{eq:sol2}
  \alpha _2  =  \frac{\alpha _3 m_3
  (v+1)^3}{m_1},
   \qquad  m_1=-\frac{m_2 m_3 (v+1)^2}{m_3 v^2+m_2}.
\end{equation}
As in the previous case the second  solution is excluded because all masses are
positive.  
With  the first solution from~\eqref{eq:al1} we get 
\begin{equation}
  \label{eq:a1so}
  \alpha _1=\frac{\alpha _3 m_3 v^3}{m_2},
\end{equation}
which together with  first solution from ~\eqref{eq:sol2} 
give the necessary condition for integrability:
  if system is integrable, then parameters $(\gamma_1, \gamma_2)$ belong 
to the curve $\Gamma_2$ given  by  
\begin{equation}
  \label{eq:Gamma2}
  \gamma_1= (1+ v)^3, \qquad \gamma_2 = v^3, \qquad v>0.
\end{equation}
Finally, if the system is integrable then parameters  $( \gamma_1, \gamma_2)$
belong to the intersection of curves $\Gamma_1$ and $\Gamma_2$ which lye in the
first quadrant of the $( \gamma_1, \gamma_2)$ plane. However, this intersection
is empty, see Fig.~\ref{intersect}.  
\begin{figure}[ht]
 \begin{center}
   \includegraphics[width=0.65\linewidth]{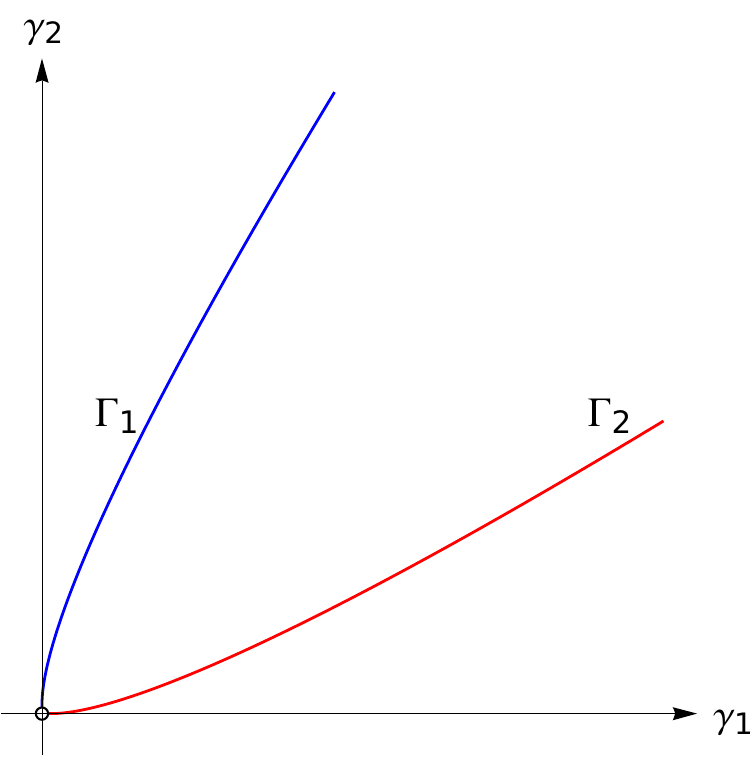}
 \end{center}
\caption{Two curves $\Gamma_1$ and $\Gamma_2$  in the $(\gamma_1,\gamma_2)$ plane of parameters}\label{intersect}
\end{figure}
\end{proof}
\begin{lemma}
\label{lem:l} If $\alpha_1=0$ and $\alpha_2\alpha_3\neq 0$, then the charged three-body problem is not integrable in the Liouville sense. 
\end{lemma}
\begin{proof}
We put   $\alpha_1=0$ and introduce variables $x$, $y$ and $u$ as in the first
step of the previous lemma proof however now these variables are generally
complex.  Making simplifications of expressions containing radicals we fix
signs of real parts of $x$ and $y$. The final conclusion does not depend on
this choice. 
\end{proof}

In \citep{Combot497649} was proven, among other things, 
the nonintegrability of the planar $n$-body problem for
arbitrary $n>2$ and arbitrary positive masses. He applied also 
Theorem~\ref{thm:mo} and Theorem~\ref{thm:mo2} choosing as a particular solution
the Euler-Moulton collinear central configuration. In his proof,  the crucial
point was to show that in general case $n>2$ matrix $\vC$ has an eigenvalue
$\lambda>1$. But for the positive masses it was proven in  Theorem 3.1 in
\cite{Pacella:87::}. Unfortunately, we do not know how to generalize it for the
case of charged bodies.

\section*{Acknowledgements}
  For AJM this research has been supported by The National Science Center of Poland under Grant No. 2020/39/D/ST1/01632 and the work of MP  was partially founded by a program of the Polish Ministry of Science under the title ‘Regional Excellence Initiative’, project no. RID/SP/0050/2024/1. For the purpose of Open Access,
the authors have applied a CC-BY public copyright license to any Author Accepted Manuscript (AAM) version arising from this submission.

\section*{Data Availability}

Data sharing is not applicable to this article, as no new data were created or analyzed in this study.

 \bibliographystyle{plainnat}

  \def\cprime{$'$} \def\cydot{\leavevmode\raise.4ex\hbox{.}} \def\cprime{$'$} \def\polhk#1{\setbox0=\hbox{#1}{\ooalign{\hidewidth \lower1.5ex\hbox{`}\hidewidth\crcr\unhbox0}}}

\end{document}